\documentclass[12pt,tightenlines,eqsecnum,floats,aps,amsmath,amssymb,nofootinbib,prd,showpacs]{revtex4}

\usepackage{amsmath,amssymb,amsfonts,amsthm}
\usepackage{graphicx}

\newcommand{\be}{\begin{equation}}
\newcommand{\ee}{\end{equation}}
\newcommand{\beq}{\begin{eqnarray}}
\newcommand{\eeq}{\end{eqnarray}}

\begin{document}

\title{Restricting Fourth Order Gravity via Cosmology}

\author{William Nelson\footnote{nelson@gravity.psu.edu}}

\affiliation{Institute for
  Gravitation and the Cosmos, The Pennsylvania State University, University Park, PA
  16801, U.S.A. }

\begin{abstract}
The cosmology of general fourth order corrections to Einstein gravity is
considered, both for a homogeneous and isotropic background and for general
tensor perturbations. It is explicitly shown how the standard cosmological
history can be (approximately) reproduced and under what condition 
the evolution of the tensor modes remain (approximately) unchanged. 
Requiring that the deviations from General Relativity are small during inflation sharpens
the current constraints on such corrections terms by some {\it thirty orders of magnitude}.
Taking a more conservative approach and requiring only that cosmology
be approximately that of GR during Big Bang Nucleosynthesis, the constraints
are improved by $4\-- 6$ orders of magnitude.

\end{abstract}


\maketitle

\section{Introduction}\label{sec:1}

General Relativity (GR) is often considered to be an extremely well tested physical theory, however compared to the
other fundamental constants (e.g. the electron charge, the speed of light etc.), the gravitational constants are
 rather poorly
constrained.  Whilst
it is true that the predictions of GR have been verified to staggering accuracy (see for example~\cite{C_Will}),
 there is little
experimental/ observational evidence to restrict corrections of GR that involve higher powers of curvature
invariants. This is simply a consequence of the fact that GR has been directly tested in weak curvature regimes,
with strong curvature tests (such as near black holes and neutrons stars) typically resulting in 
theoretical restrictions.
The Einstein-Hilbert action for GR contains only terms that are, at most, second order in the derivatives of
the metric (in the form of the Ricci scalar $R$) and a natural question to then ask is what restrictions are there
on the presence of terms that contain higher order derivatives of the metric?

Motivation for such a question can be found from two, broadly different, approaches. The first is phenomenological:
are there corrections to the Einstein-Hilbert action that can better describe observed data? In particular
there has been much effort in looking for deviations from GR that might explain the apparent presence of
Dark Energy, Dark Matter and inflation
(such as $f(R)$ gravity~\cite{Nojiri:2006ri,Appleby:2007vb,Sotiriou:2008rp,Nojiri:2003ft,Amendola:2006we,
Nojiri:2008nt,Nojiri:2010wj,Capozziello:2007ec,Capozziello:2006uv,Nojiri:2007as,Boehmer:2007kx},
 chameleon models~\cite{Khoury:2003aq,Khoury:2003rn,Brax:2008hh}, conformal gravity~\cite{Mannheim:1999nc,Mannheim:1996rv},
 MOND~\cite{Milgrom:1992hr,Milgrom:2002tu,Bekenstein:2007iq},
TeVeS~\cite{Skordis:2009bf,Moffat:2005si,Sagi:2007hb,Skordis:2005eu,Sanders:2005vd,Zlosnik:2006sb} etc.).
 The second
approach is to consider theoretically motivated corrections to GR, often due to Quantum Gravity (such as
String/ Brane theory~\cite{Mavromatos:2007sp,Smolin:1995ai,Baumann:2009ni,Brax:2003fv,Duff:2000mt,Garriga:1999yh,
Nelson:2008sv,Germani:2002pt,Hawking:2000kj,Kachru:2003sx,Randall:1999vf,Sasaki:1999mi},
ADS-CFT~\cite{Anchordoqui:2000du,Gubser:1999vj,Nojiri:2000kq},
 Loop Quantum Gravity~\cite{Bojowald:2006ww,lving_rev_MB,Smolin:2010kk, Nelson:2007cp,
Bianchi:2009ri,Bianchi:2006uf,Smolin:2004sx,Ashtekar:2004eh,Thiemann:2007zz,Baez:1999sr},
 Chern-Simons theory~\cite{Alexander:2009tp,Yunes:2008ua}
etc.)
or a Grand Unification theory (such as Non-commutative
Geometry~\cite{Buck:2010sv,Chamseddine:2010ud,Chamseddine:2006ep}). Because there are only
three curvature invariants that contain at most two derivatives of the metric, $R$, the cosmological
constant $\Lambda$ and the Gauss-Bonnet combination (defined below), one is generally forced to consider
correction terms containing $4^{\rm th}$ and higher derivatives of the metric\footnote{If these terms
are built out of curvature invariants there is always an even number of derivatives.}, at least in the
low energy, effective limit of the full theory.

Theoretically there is, however, a serious difficultly with a theory that contains higher than second
order derivatives of the metric; such a theory would contain `ghosts'. These are modes that have a negative kinetic
term and can result in super-luminal propagation and other instabilities~\cite{Barth:1983hb,Stelle:1977ry,
DeFelice:2006pg}.
Such difficulties become particularly acute when the theory is quantised, since the
presence of ghosts leads to particles with negative energies or states with negative norm.
This shows that any theory containing higher order corrections
to GR cannot be a fundamental
theory unless powerful non-perturbative effects come into play. Here we will consider a general
$4^{\rm th}$ order correction to GR and take the point of view that this is an {\it effective} theory,
approximating (some of) the corrections to GR that are produced by some complete underlying, non-perturbative
theory. This precisely is the approach that is taken when renormalisation group techniques are applied to
GR~\cite{Lauscher:2001ya,Reuter:2001ag,Donoghue:1993eb}.

The only terms containing at most $4^{\rm th}$ order derivatives of the metric that are constructed
out of curvature invariants (other than the cosmological constant) are
\be\label{eq:curv_inv}
 R^{\mu\nu\rho\gamma}R_{\mu\nu\rho\gamma}~,~~ R^{\mu\nu}R_{\mu\nu}~,~~ R^2~,~~ R~,
\ee
where $R^{\mu}_{~\nu\rho\gamma}$ is the Riemann tensor, defined as
\be
R^{\mu}_{~\nu\rho\beta} = \Gamma^{\mu}_{\rho\nu,\beta} - \Gamma^{\mu}_{\beta\nu,\rho} + \Gamma^{\alpha}_{\nu\rho}
\Gamma^{\mu}_{\beta\alpha} - \Gamma^{\alpha}_{\nu\beta} \Gamma^{\mu}_{\rho\alpha}~,
\ee
and we follow the notation of~\cite{Nelson:2010ig} Greek letters to denote space-time indices; $\Gamma^{\mu}_{\nu\rho}$ is the usual
Christoffel symbol; we denote partial derivatives as $\partial_\eta R = R_{,0}$ and
we use the convention $R_{\mu\nu} = R^\rho_{~\mu\rho\nu}$, with the signature $\left( - + + +\right)$.
The Ricci scalar (last term in (\ref{eq:curv_inv})) is at most second order in derivatives of the
metric, while the particular combination, $R^2 - 4R_{\mu\nu}R^{\mu\nu}+R^{\mu\nu\rho\gamma}R_{\mu\nu\rho\gamma}$,
called the Gauss-Bonnet combination, satisfies (in four dimensions) 
\be\label{eq:BG}
 \sqrt{-g}\left( R^2 - 4R_{\mu\nu}R^{\mu\nu} + R^{\mu\nu\rho\gamma}R_{\mu\nu\rho\gamma} \right) = {\rm total~
divergence}~.
\ee
Thus, assuming there are no boundary terms (which can be a significant complication, particularly when 
attempting to quantise the theory~\cite{Barth:1983hb}), one can write any $4^{\rm th}$ derivative correction to the Lagrangian
of GR as
\be\label{eq:action}
 {\cal S} = -\int {\rm d}^4 x \sqrt{-g} \left[ \frac{\gamma}{\kappa^2} R - \hbar \beta R^2 + \hbar \alpha R_{\mu\nu}R^{\mu\nu}\right]
+{\cal S}_{\rm matter}~,
\ee
where for simplicity we have neglected the cosmological constant. Note that higher order terms (e.g. $R^3$) can lead
to $4^{\rm th}$ derivative terms in the equation of motion~\cite{Schmidt:2001ac}, however here we will restrict our attention to terms that are
of similar order in the Lagrangian, assuming that Eq.~(\ref{eq:action}) is a low curvature expansion of some underlying theory.
We define $\kappa^2 = 32\pi G$ where $G$
is Newton's constant and we use units in which $\left[c\right]=1$. 
$\alpha$, $\beta$ and $\gamma$ are dimensionless couplings and the presence of $\hbar$ is due only to the dimensions (i.e.\ the 
theory is entirely classical). In general
the coefficients $\alpha$ and $\beta$ are arbitrary constants, however if these corrections are due to some underlying
Quantum Gravity theory, then they would be expected to be of order one, so that the quantum gravity scale is of the order
of the Planck scale. The value of $\alpha$ and $\beta$ set the scale at
which significant corrections to the GR occur and in the following we will restrict these values via cosmological considerations.
In all of the following we set  $\left[ \hbar\right]=1$.

By varying this action with respect to the metric $g_{\mu\nu}$ as usual, one finds the field equations~\cite{Stelle:1977ry},
\beq\label{eq:EoM}
 H^\mu_{~\nu} &\equiv& \left( \alpha - 2 \beta\right) R^{;\mu}_{~~;\nu} - \alpha R^{\mu~;\rho}_{~\nu~;\rho} - \left( \frac{\alpha}{2} 
- 2 \beta \right) g^\mu_{~\nu} R^{;\rho}_{~;\rho} + 2\alpha R^{\rho\lambda} R^\mu_{~\rho \nu\lambda} \nonumber \\
&& - 2\beta R R^\mu_{~\nu} - \frac{1}{2} g^\mu_{~\nu} \left( \alpha R^{\rho \lambda}R_{\rho\lambda} - \beta R^2\right)
+ \gamma \kappa^{-2} G^\mu_{~\nu} = -\frac{1}{2} T^\mu_{~\nu}~,
\eeq
where $R_{;\mu}=\nabla_\mu R$ is the covariant derivative of $R$ and $T_{\mu\nu}$ is the energy momentum tensor
given by varying ${\cal S}_{\rm matter}$ with respect to $g_{\mu\nu}$. Perturbations of this theory around a flat
background, show that the theory contains, in addition to 
the massless graviton, a massive spin-$2$ field (of negative energy) which is a ghost field and a (positive energy)
massive scalar field~\cite{Stelle:1977ry}.

One particular example of a theory that predicts $4^{\rm th}$ order corrections to GR is Non-commutative 
Geometry~\cite{Chamseddine:2010ud}.
Here the asymptotic expansion of this (non-commutative) geometric theory produces the entire
standard model coupled to a gravitational action of the form~\cite{Chamseddine:2006ep},
\be
 {\cal S}_{\rm NCG} = - \int {\rm d}^4 x \sqrt{-g} \left( \frac{1}{16\pi G}R + \alpha_{\rm NCG}C^{\mu\nu\rho\gamma}C_{\mu\nu\rho\gamma}\right)~,
\ee
where $C_{\mu\nu\rho\gamma}$ is the Weyl tensor\footnote{It should be mentioned that this Non-commutative Geometry theory
is formulated in a Euclidean signature and is entirely classical.}. The consequences of this modification for 
cosmology~\cite{Nelson:2009wr,Nelson:2008uy}
and astrophysics~\cite{Nelson:2010rt,Nelson:2010ru} have been considered and in particular in~\cite{Nelson:2008uy}
 it was shown that
background cosmology is unaffected by the presence of such a correction. Using (\ref{eq:BG}) to write 
action this in the form of (\ref{eq:action}) gives,
\be
  {\cal S}_{\rm NCG} = - \int {\rm d}^4 x \sqrt{-g}  \left[ \frac{1}{16\pi G} R +2\alpha_{\rm NCG}\left(
   R_{\mu\nu}R^{\mu\nu}-\frac{1}{3}R^2\right)\right]~.
\ee
In particular then
corrections of this form satisfy $\alpha = 3\beta$ and in the following we will show that
{\it only} this combination leaves the background dynamics unaffected.

The question we will address is what restrictions on $\alpha$ and $\beta$ can be deduced from the 
cosmology given by (\ref{eq:action}) and how they compare to the current constraints.
Existing restrictions on $\alpha$ and $\beta$ are in fact very mild, requiring only that these
 coefficients be less than ${\cal O}\left( 10^{72}\right)$.
The restrictions come from solar system tests, notable the perihelion precession of
Mercury~\cite{Stelle:1977ry} and gravitational wave production in binary systems~\cite{Nelson:2010rt}.
There are also laboratory scale tests of the inverse square law, which restrict parameters of
this form significantly more, $\alpha < {\cal O} \left( 10^{60}\right)$~\cite{Kapner:2006si}.
However here we will focus on the cosmological implications of such corrections and so compare to
the other large scale constraints.

The reason that these restrictions are so weak is
that corrections to General Relativity occur at higher orders in curvature and hence are highly suppressed in
weak curvature systems. In order for the effects of such corrections to become more significant,
one needs to consider
a system that contains either strong curvature (such as black holes or the early universe) or long evolution times (so as
to allow the small deviations from General Relativity to accumulate). Here we consider the latter case, by examining
both the dynamics of the background cosmology and 
the evolution of tensor mode perturbations over cosmological time scales.

Ideally one would want to consider the evolution of scalar mode perturbations, since these can be directly related
to the wealth of observational data coming from large scale galaxy surveys, weak lensing maps and the Cosmic Microwave
Background (CMB). By contrast, cosmological tensor mode perturbations, although generically predicted by inflationary
models, have yet to be observed. Despite this, tensor mode perturbations have the advantage of being technically
much simpler to calculate than scalar modes (evolving according to a single equation, rather than four coupled 
equations).
 In addition, since tensor perturbations do not couple to the matter content of the universe other than
through the background evolution (at least for matter with vanishing anisotropic stress)
 they are a direct probe of the underlying gravity theory and are unaffected
by any possible modifications to the matter action.

In Section~\ref{sec:BG} we derive the dynamics of the background (i.e.\ homogeneous and isotropic)
cosmology, discussing in particular the solutions during inflation and during the radiation and
matter dominated eras. Section~\ref{sec:tensors} derives the general evolution equation for tensor mode
perturbations for the action given in (\ref{eq:action}), while in Section~\ref{sec:tensors_BG}
we consider the evolution of such perturbations against specific background cosmologies.
Using these evolution equations we derive constraints on $\alpha$ and $\beta$ in Section~\ref{sec:constraints}
and summarise and conclude in Section~\ref{sec:conc}.

\section{Background Cosmology}\label{sec:BG}
In order to examine the background cosmology of this theory, we need to calculate the modified
Friedmann and Raychaudhuri equations, which are given respectively by, $H^0_{~0} = \frac{-1}{2} T^0_{~0}$ 
and $H^i_{~i} = \frac{-1}{2} T^i_{~i}$.
We want to consider a Friedmann-Robertson-Walker universe, whose metric is given in block
diagonal form by,
\be\label{eq:metric_bg}
 \bar{g}_{\mu\nu} = a^2\left(\eta\right) \left( \begin{array}{cc}
                                                    -1 & {\bf 0}^{\top} \\
                                                    {\bf 0} & \gamma_{ij} \end{array} \right)~,
\ee
where $i,j\dots=1,2,3$ will be used to denote spatial components; ${\bf 0}$ is the three dimensional
zero vector; $a(\eta)$ is the scale factor, with $\eta$ conformal time, and the three dimensional
spatial metric $\gamma_{ij}$ is given by,
\be
 \gamma_{ij} = \frac{\delta_{ij} }{\left( 1+\frac{K}{4} \delta_{mn}x^m x^n\right)^2}~,
\ee
where $\delta_{ij}$ denotes the three dimensional Kronecker delta and
$K$ the (scaled) spatial curvature ($K=1$ has spherical spatial topology and corresponds to a closed universe,
$K=0$ is spatially flat and $K=-1$ is spatially hyperbolic).
We shall denote background quantities with a bar. For example, the full metric is given by,
\be
 g_{\mu\nu}\left( \eta, {\bf x}\right)
 = \bar{g}_{\mu\nu}\left( \eta\right) + \delta g_{\mu\nu} \left( \eta, {\bf x} \right)~,
\ee
where, as expected from homogeneity and isotropy, the background quantity (in this case the metric)
depends only on conformal time, whilst the perturbed quantity (here $\delta g_{\mu\nu}$) depends on
both spatial and temporal coordinates.

With this metric one can immediately calculate the components of the curvature tensors. For later convenience
we list some of the non-zero components the of Riemann and Ricci tensors and the Ricci scalar, with the remaining
non-zero components being given directly from the symmetries of the tensors.
\beq\label{eq:curvs}
 &&\bar{R}^0_{~i0j} = -{\cal H}' \gamma_{ij}~, ~~~~~ \bar{R}^i_{~00j} = -{\cal H}'\delta^i_j~,\nonumber \\
\nonumber \\
&& \bar{R}^i_{~jkm} = -\left( {\cal H}^2 +K\right)\left( \delta^i_k\gamma_{jm} - \delta^i_m\gamma_{jk}\right)~,~~~~
 \bar{R}^0_{~0} = -\frac{3}{a^2}{\cal H}'~,\nonumber \\ 
&&  \bar{R}^i_{~j} = -\frac{1}{a^2} \left( {\cal H}'
+2\left( {\cal H}^2 + K\right)\right)\delta^i_j~, ~~~~
 \bar{R} = \frac{-6}{a^2} \left( {\cal H}' + {\cal H}^2 + K\right)~,
\eeq
where a dash means differentiation with respect to conformal time $\eta$ and ${\cal H} \equiv a'/a$ is the
{\it conformal} Hubble parameter (see for example~\cite{durrer_book}).

As outlined in Appendix~A, one can use this metric to find,
\beq\label{eq:Fried_ray_BG}
 \bar{H}^0_{~0} &\equiv& \frac{2}{a^2} \frac{\left( \alpha - 3\beta\right)}{\gamma\kappa^{-2}} 
\left[ -2{\cal H} \left( {\cal H}'' + 2{\cal H}
{\cal H}'\right) + 3{\cal H}^4 + 2{\cal H}^2K + \left( {\cal H}'\right)^2 -K^2\right] \nonumber \\
&& + \left( {\cal H}^2+K\right) = \frac{1}{6\gamma\kappa^{-2}}a^2\rho~, \\
\label{eq:Ray_ray_BG}
 \bar{H}^{\rho}_{~\rho} &=& \frac{-2}{a^2} \frac{\left( \alpha-3\beta\right)}{\gamma\kappa^{-2}}
\left[ {\cal H}''' - 6{\cal H}^2{\cal H}' - 2{\cal H}'K\right] \nonumber \\
&& + \left( {\cal H}' +{\cal H}^2
+K\right) = \frac{1}{12\gamma\kappa^{-2}} a^2\left( \rho - 3P\right)~,
\eeq
where we used the fact that by homogeneity and isotropy the most general
background energy momentum tensor is that of a perfect fluid,
\be
 \bar{T}_{\mu\nu}\left(\eta\right) = \left( \begin{array}{cc} -\bar{g}_{00} \rho\left( \eta\right) & {\bf 0}^\top \\
                                         {\bf 0} & \bar{g}_{ij} P\left(\eta\right) \end{array} \right)~,
\ee
with $\rho$ and $P$ the energy density and pressure of the fluid.

Notice that, as expected, the $\alpha=\beta=0$ case i.e.\ GR, gives the usual Friedmann and Raychaudhuri equations,
as does $\alpha=3\beta$. Recall that latter case corresponds to the NCG motivated correction to GR (see Section~\ref{sec:1}),
which vanishes for a FRW universe~\cite{Nelson:2008uy}.

Unfortunately there does not appear to be a general, analytic solution to (\ref{eq:Fried_ray_BG})
for a fluid satisfying $P=\omega \rho$, with $\omega$ a constant, however progress can be made by
checking the consistency of certain, important, cosmological solutions. In particular we will consider radiation and 
matter dominated universes, whose matter components have  $\omega=1/3$ and $\omega =0$ respectively
and also a universe dominated by a slowly rolling scalar field\footnote{Recall that the potential
energy and pressure associated with a scalar field evolving in a potential $V\left(\phi\right)$ are
$\rho_\phi=\dot{\phi}^2/2 + V\left( \phi\right)$ and $P_\phi = \dot{\phi}^2/2 - V\left( \phi\right)$ respectively.
Thus if the field is slowly rolling i.e.\ if  $V\left( \phi\right) \gg \dot{\phi}^2/2$ then the
pressure is given by $P_\phi \approx - \rho_\phi$
}, for which $\omega = -1$. In~\cite{Macrae:1981ic}
the conformal properties of (\ref{eq:action}) were used to derive general consequences of these corrections, however here
we explicitly calculate the background evolution, to facilitate the inclusion of tensor perturbations
in Section~\ref{sec:tensors}.

One would like
to be able to specify the form of the matter content (i.e.\ $\omega$) and deduce the corresponding cosmological
history $a\left(\eta\right)$, however this leads to rather involved expressions and a more transparent
approach is to specify the form of the cosmological expansion $a\left(\eta\right)$ that matches the 
GR expectation and evaluate the properties of the matter components that support this expansion.
The properties of these matter fluids will, in general, deviate from their standard values
and this deviation measures the significance of the correction terms on the cosmological evolution.
It is important to note here that the non-linear nature of Eq.~(\ref{eq:Fried_ray_BG}) and Eq.~(\ref{eq:Ray_ray_BG})
means that finding such deviations to be small, does not immediately imply that the solutions, $a\left(\eta\right)$,
with the correct matter content would be close to those of standard GR. The approach used here is however
a necessary (but not sufficient) condition for the cosmological evolution to approximate that of GR and should
be supplemented by direct numerical evaluations of  Eq.~(\ref{eq:Fried_ray_BG}) and Eq.~(\ref{eq:Ray_ray_BG}).

\subsection{De Sitter expansion}\label{sec:ds_bg}
The scale factor for an exponentially expanding universe (in conformal time) is given by
\be
a\left(\eta\right) =\frac{a_{\rm inf}\eta_{\rm inf}}{\eta}~,
\ee
where $a_{\rm inf}$ is the normalisation of the scale factor at the beginning of inflation,
taken to occur at (conformal) time $\eta_{\rm inf}$.
Using this ansatz in (\ref{eq:Fried_ray_BG}) and (\ref{eq:Ray_ray_BG}) we find,
\beq\label{eq:inf_eqns}
 \rho\left( a\right) &=& \frac{12\left( \alpha-3\beta\right)}{\left(a_{\rm inf}\eta_{\rm inf}\right)^4}
 \left( -4+2K\left(\frac{a_{\rm inf}\eta_{\rm inf}}{a}\right)^2 
 - K^2\left( \frac{a_{\rm inf}\eta_{\rm inf}}{a}\right)^4\right) \nonumber \\
&& + \frac{6\gamma \kappa^{-2}}{\left(a_{\rm inf}\eta_{\rm inf}\right)^2} \left( 1 + K
\left(\frac{a_{\rm inf}\eta_{\rm inf}}{a}\right)^2\right)~, \nonumber \\
 \rho\left( a\right) \left(1-3\omega\right) &=& \frac{24\left(\alpha-3\beta\right)K}{\left(a_{\rm inf}
\eta_{\rm inf}\right)^2 a^2}  + \frac{ 12\gamma\kappa^{-2}}{\left(a_{\rm inf}\eta_{\rm inf}\right)^2}
 \left( 2+K\left(\frac{a_{\rm inf}\eta_{\rm inf}}{a}\right)^2\right)~.
\eeq
In a spatially flat universe ($K=0$),  we then have
\be\label{eq:inf_rho}
 \rho\left( a\right) = \frac{6\gamma \kappa^{-2}}{\left(a_{\rm inf}\eta_{\rm inf}\right)^2}
  \left( 1- 8\frac{A}{\left(a_{\rm inf}\eta_{\rm inf}\right)^2} \right)~,
\ee
and
\be\label{eq:inf_omega}
 \omega = \frac{1}{3} \left( 1 - 4\left( 1-8\frac{A}{\left(a_{\rm inf}\eta_{\rm inf}\right)^2} \right)^{-1}
\right)~,
\ee
where
\be\label{eq:A}
 A=\frac{\alpha-3\beta}{\gamma\kappa^{-2}}~.
\ee
Recall that both $\kappa$ and {\it cosmic} time $t\equiv a\eta$ have units of length (or equivalently $M^{-1}$),
hence, on dimensional grounds, one expects the key {\it dimensionless} parameter to be $A\left( a\eta\right)^{-2}$.

\begin{figure}
 \begin{center}
  \includegraphics[scale=1.0]{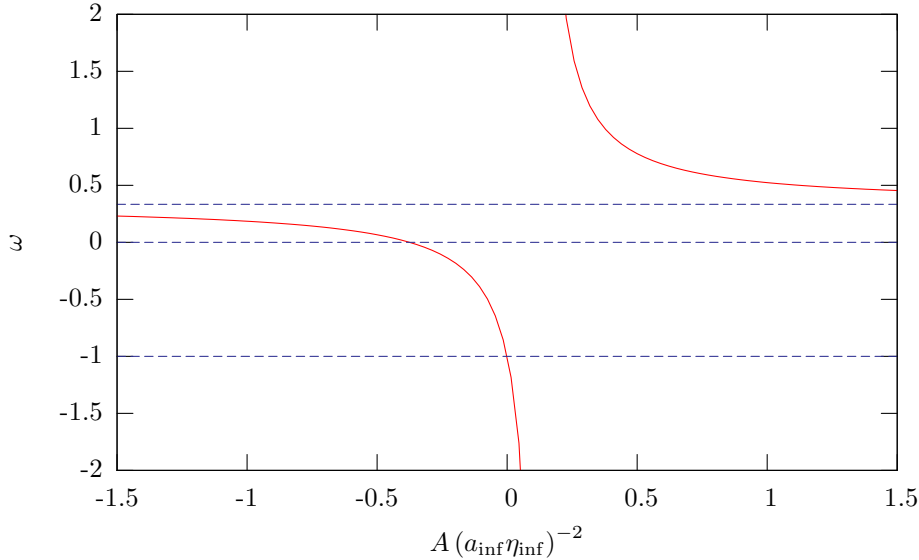}
   \caption{\label{fig:1} The equation of state that is required to maintain an exponential expansion depends on
    the coefficients of the corrections to GR, $A\left( a_{\rm inf}\eta_{\rm inf}\right)^{-2}$. Also plotted for
    comparison are the equations of state for a cosmological constant $\omega = -1$, pressureless matter
    $\omega = 0$ and radiation $\omega=1/3$.}
 \end{center}
\end{figure}

From (\ref{eq:inf_rho})
we see that exponential expansion is given by a fluid with a constant energy density just as for GR,
however for a given energy density the Hubble rate, ${\cal H}$, will depend on the parameters of the theory
($\alpha$, $\beta$ and $\gamma$).
In order to maintain the exponential expansion the equation of state parameter of the fluid,
$\omega$ differs from the GR case ($\omega = -1$). Although it is a constant, it is important to 
note that this is only true for exactly exponential expansion with $K=0$, in general one would
require a varying $\omega$. The equation of state that is required depends on the
$\alpha$, $\beta$ and $\gamma$ and is plotted in Fig.~(\ref{fig:1}), notice in particular
for $8A=\left(a_{\rm inf}\eta_{\rm inf}\right)^2$ the equation of state that produces exponential
expansion diverges, essentially meaning that it is not possible to have such expansion for these for 
a fluid with a finite pressure. However for $8A=\left(a_{\rm inf}\eta_{\rm inf}\right)^2$,
(\ref{eq:inf_rho}) gives $\rho\left(a\right)=0$, thus it is not surprising that the pressure becomes 
ill defined. For this specific choice of parameters, the vacuum solution of the theory
is an exponentially expanding cosmology. This divergence will appear again (for the $\alpha=0$ case)
in Section~\ref{sec:tensors}, when perturbations are considered.

From Fig.~(\ref{fig:1}) we see that for $0 < 8A\left( a_{\rm inf}\eta_{\rm inf}\right)^{-2} < 1$
 a phantom fluid is required to produce the expansion i.e.\ $\omega < -1$, 
while for $8A\left( a_{\rm inf}\eta_{\rm inf}\right)^{-2}>1$
 a fluid with $\omega >1/3$ is necessary. Finally for $A<0$ (i.e.\ $\alpha - 3\beta <0$)
exponential expansion can be achieved by fluids with $-1<\omega <1/3$. In particular, only $A=0$ 
(or equivalently $\alpha-3\beta=0$) will produce an exponential expansion in the presence of a
slowly rolling scalar field (for which $\omega \approx -1$). The $A<0$ and $8A\left( a_{\rm inf}\eta_{\rm inf}\right)^{-2}>1$
cases could have the
desired expansion driven by a scalar field if the kinetic and potential energies of the field were correctly
matched, to give (\ref{eq:inf_omega}). However even if this
does not result in excessive fine-tuning (over that required by standard inflation), it would likely be
rather difficult to ensure that $\omega$ has the correct value over a long period of expansion.

Thus a period of exponential expansion, within general fourth derivative gravity theories is
unlikely to be sourced by a simple scalar field, unless $A\ll 1$ (this is the case for example in~\cite{Muller:1988db} and
\cite{Schmidt:1988hq}).
However the presence of non-minimal couplings
between the scalar field and the curvature terms can have a significant effect even in the
$\alpha - 3\beta=0$ case~\cite{Nelson:2009wr,Buck:2010sv,Barvinsky:2009ii},
so one cannot rule out the possibility. Indeed, the motivation for
studying such corrections to GR requires that we consider $\alpha-3\beta$ be small (and hence $A$ to be small), in order
to avoid super-luminal propagation (and other ghost effects, see Section~\ref{sec:1}). With this in mind it is
possible that standard slow roll inflation would result in {\it almost} exponential expansion
i.e.\ that the deviations from the expected GR behaviour are small. Determining the observational
consistency of
such an approximation would require both a general solution to (\ref{eq:Fried_ray_BG})
in the presence of a scalar field and, eventually, details of the evolution of perturbations
against this background.

If we want to have inflation driven by a scalar field in the standard way, the correction
terms need to satisfy, $|A|\left( a_{\rm inf}\eta_{\rm inf}\right)^{-2}\ll 1$. Since this
is a constant, it would imply that also at the end of the inflationary epoch
$|A|\left( a_{\rm inf}\eta_{\rm inf}\right)^{-2}\ll 1$ and
hence the radiation era would begin with this dimensionless ratio being small.

\subsection{Radiation era}
In standard GR an era dominated by a single fluid with $\omega=1/3$ (i.e.\ radiation)
 produces $a\left( \eta\right)=a_{\rm r}\eta/\eta_{\rm r}$, where $a_{\rm r}$ is the scale 
factor at $\eta_{\rm r}$, which is the conformal time at which the radiation era begins.
Using this form of the scale factor in (\ref{eq:Fried_ray_BG}), we find that
\beq\label{eq:rad_eqn1}
 \rho\left( a\right) &=& \frac{12\left( \alpha - 3\beta\right) a_{\rm r}^4}{
\eta_{\rm r}^4a^8}\left( 4+ 2K\left(\frac{a\eta_{\rm r}}{a_{\rm r}}\right)^2
- K^2\left( \frac{a\eta_{\rm r}}{a_{\rm r}}\right)^4\right) \nonumber \\
&& + \frac{\gamma\kappa^{-2}a_{\rm r}^2}{\eta_{\rm r}^2a^4} \left( 1 + K \left(
\frac{a\eta_{\rm r}}{a_{\rm r}}\right)^2\right)~,
\eeq
and
\beq\label{eq:rad_eqn2}
 \rho\left( a\right) \left( 1-3\omega\right) &=& -\frac{48\left( \alpha-3\beta\right)K a_{\rm r}^2}{
\eta_{\rm r}^2 a^6} +  \frac{12\gamma\kappa^{-2} K}{a^2}~.
\eeq
Thus, if we restrict our attention to the $K=0$ case, we find that an expansion of $a\left( \eta\right)
=a_{\rm r}\eta/\eta_{\rm r}$ is produced by a fluid with $\omega = 1/3$ and
\be\label{eq:rad}
 \rho\left( a\right) = \frac{6a_{\rm r}^2 \gamma\kappa^{-2}}{\eta_{\rm r}^2a^4} \left( 1+ 
 8\frac{Aa_{\rm r}^2}{\eta_{\rm r}^2a^4}\right)~,
\ee
where $A$ is given in (\ref{eq:A}).
Such a fluid has the equation of state of radiation, however the required energy density
would be larger (for $\alpha-3\beta >0$) than that of radiation and would scale differently.
Such corrections to the energy density would, however, rapidly become small as the universe expands and
\be\label{eq:rad_con}
 8\frac{|A|a_{\rm r}^2}{\eta_{\rm r}^2a^4} \ll 1~.
\ee

As discussed in Section~(\ref{sec:ds_bg}),
if the radiation era is preceded by an epoch of inflation that is driven by a scalar
field, then $8|A|\ll \left( a_{\rm r}\eta_{\rm r}\right)^2$
at the beginning of the radiation era. This would ensure
that the expected expansion would approximately be achieved by standard radiation
even early in the radiation era. 

\subsection{Matter era}
Repeating the above procedure for matter era type expansion, i.e., $a\left(\eta\right)
=a_{\rm m}\eta^2/\eta_{\rm m}^2$, where as in the previous two cases,
we have defined the time the matter era began as $\eta_{\rm m}$ and the scale factor at
this time to be $a_{\rm m}$. One then finds (again for the $K=0$ case) that the required
energy density is,
\be
\rho\left( a\right) = \frac{24 \gamma \kappa^{-2}a_{\rm m}}{\eta_{\rm m}^2a^3}\left( 1 + 
34\frac{Aa_{\rm m}}{\eta_{\rm m}^2a^3}\right)~.
\ee
Thus, just as in the radiation case, the fluid would have to have a larger (for 
$\alpha - 3\beta >0$) energy density than standard pressureless matter. These
corrections would rapidly become very small as the universe expands, however,
unlike the radiation era case where the usual equation of state was found, here
we require,
\be
 \omega = \frac{2Aa_{\rm m}}{\eta_{\rm m}^2a^3}\left( 1+34\frac{Aa_{\rm m}}{\eta_{\rm m}^2a^3}
\right)^{-1}~.
\ee
Therefore an exactly pressureless fluid will not produce an expansion consistent with predictions
of GR. However the pressure that is required to match the GR expectations is
vanishingly small for 
\be
 \frac{|A|a_{\rm m}}{\eta_{\rm m}^2a^3} \ll 1~.
\ee

Thus for large scales (or late times) the background cosmological evolution
will be approximately the same as that expected from GR, provided
\be
 |A|\left(a_{\rm inf}\eta_{\rm inf}\right)^{-2}\ll 1~.
\ee
Considering for the moment only the radiation and matter eras, all that is required is that $|A|\left(
a_{\rm r}\eta_{\rm r}\right)^{-2}$ be
sufficiently small, so as to allow (\ref{eq:rad}) to be approximately
correct for standard radiation. As the universe expands the evolution will then quickly
approach the standard cosmology of GR. For $|A|\left(a_{\rm inf}\eta_{\inf}\right)^{-2}\ll 1$
this is satisfied and hence the standard background dynamics can be (approximately) reproduced
from a slowly rolling scalar field, with the corrections to the GR expectations remaining small
throughout the history of the universe. This agrees with the conclusions of~\cite{Macrae:1981ic}.

We have considered only the flat ($K=0$) case, the presence of a non-zero curvature would
introduce additional corrections to the standard GR evolution equations. However it is clear from
(\ref{eq:inf_eqns}), (\ref{eq:rad_eqn1}) and (\ref{eq:rad_eqn2}) (and similarly
for the analogous equations for the matter era) that the consequences of $K\neq 0$ rapidly
become small in a large scale, expanding universe. This is to be expected since the 
energy density associated with the spatial curvature
of an expanding universe decreases with time, hence its consequences will diminish.

\section{Tensor mode perturbations}\label{sec:tensors}

In the previous sections we derived the evolution of the background cosmology for a general
$4^{\rm th}$ order correction to GR. Given the prevalence of ghost modes in such theories,
it is important to examine the evolution of perturbations on these background cosmologies.
Here we derive the equations governing the dynamics of cosmological gravitational waves. See~\cite{Bogdanos:2009tn} for
an alternative derivation in a more general setting and~\cite{Noh:1996da} for an derivation similar to the one presented here.
Also~\cite{Hwang:2001pu} for the generation and evolution of tensor modes during
inflation for a specific $f(R)$ model, which includes a detailed analysis of the properties of the resulting spectrum.

A general perturbation to the metric,~(\ref{eq:metric_bg}),
can be decomposed into scalar, vector and tensor components (see for example~\cite{durrer_book})
and here we will be concerned only with the tensor modes, which are transverse and traceless and given by
the perturbed metric,
\be
 \delta g_{00} = \delta g_{0i} = 0~, ~~~~ \delta g_{ij} = 2a^2\left(\eta\right) E_{ij}~,
\ee
where the perturbation $E_{ij}$ is transverse ($D^i E_{ij}=0$) and traceless ($\bar{g}^{ij}E_{ij}=0$), 
the non-transverse and non-traceless parts giving contributions to the vector and scalar perturbations
respectively. Using this
perturbed metric one finds the perturbations to the components of the Riemann tensor,
\beq\label{eq:curvs_pert}
 && \delta R^0_{~i0j} = - \left( E''_{ij} + {\cal H}E'_{ij} + 2{\cal H}'E_{ij} \right)~,~~~~
    \delta  R^0_{~ijk} = E'_{ij|k} - E'_{ik|j}~,\nonumber \\
 && \delta R^i_{~00j} = \left( E''\right)^i_{~j} + {\cal H}\left( E'\right)^i_{~j}~, ~~~~
    \delta R^i_{~0jk} = - \left( E'\right)^i_{~k|j} + \left( E'\right) ^i_{~j|k}~ \nonumber \\
 && \delta R^i_{~j0k} =- \left( E'\right)^{~~|i}_{jk} - \left( E'\right) ^i_{~k|j}~, \nonumber \\
 && \delta R^i_{~jmn} = -2{\cal H}^2 \left( \delta^i_m E_{jn} - \delta ^i_n E_{jm}\right) \nonumber \\
&& ~~~~~~~~~~~         - \left( E^i_{~j|nm} - E^i_{~j|mn} + E^i_{~n|jm} - E^i_{~m|jn}
                             + E^{~~~|i~}_{jm~|n} - E^{~~|i~}_{jn~|m} \right)
\nonumber \\
&& ~~~~~~~~~~~         - {\cal H} \left( \delta^i_m\left( E'\right)_{jn} - \delta^i_n\left( E'\right)_{jm}
                       -\left( E'\right)^i_{~n}\gamma_{jm} + \left( E'\right)^i_{~m}\gamma_{jn}\right)~,
\eeq
where $E_{ij|k} = D_k E_{ij}$ is the spatial covariant derivative of $E_{ij}$.
Similarly one can calculate the perturbations to the remaining tensor components given in (\ref{eq:curvs}),
the only non-zero expression being,
\beq\label{eq:metric_perts}
 && \delta R^i_{~j} = \frac{1}{a^2} \left( - \left(E''\right)^i_{~j} - 2{\cal H}\left( E'\right)^i_{~j} + \left( \Delta -K\right)
  E^i_{~j}\right)~,
\eeq
where $\Delta = D^i D_i$ is the spatial Laplacian. Note also that, because $E_{ij}$ is traceless, we have $\delta g^i_{~i} = 0$,
and hence $\delta G^i_{~j} = \delta R^i_{~j}$.

With this one can calculate the perturbations to (\ref{eq:EoM}) to find,
\beq\label{eq:EoM_pert}
 \delta H^\mu_{~\nu} = \left( \alpha - 2\beta\right) \delta \left( R^{;\mu}_{~~;\nu}\right) - \alpha\delta\left( 
R^{\mu~;\rho}_{~\nu~~;\rho}\right) - \left( \frac{\alpha}{2} - 2\beta\right) \bar{g}^{\mu}_{~\nu} \delta\left(
R^{;\rho}_{~;\rho}\right) + 2\alpha\left( \delta R\right)^{\rho\lambda} \bar{R}^\mu_{~\rho\nu\lambda}\nonumber \\
 +2\alpha\bar{R}^{\rho\lambda}\left(\delta R\right)^\mu_{~\rho\nu\lambda} + \left( \gamma \kappa^{-2} - 2\beta \bar{R}
\right) \left( \delta R\right)^{\mu}_{~\nu} = \frac{-1}{2} \delta T^{\mu}_{~\nu}~.
\eeq
Of the various terms appearing in (\ref{eq:EoM_pert}) those not involving covariant derivatives are the
easiest to calculate. Using (\ref{eq:curvs_pert}) and (\ref{eq:metric_perts}), one immediately finds,
\be
 \left( \delta R\right)^{\rho\lambda}\bar{R}^i_{~\rho j \lambda} = -\frac{ \left({\cal H}^2 + K \right)}{a^4}
\left( \left( E''\right)^i_{~j} + 2{\cal H}\left( E'\right)^i_{~j} - \left( 2{\cal H}' + 4{\cal H}^2+\Delta +2 K\right)E^i_{~j}\right)~,
\ee
and
\beq
\bar{R}^{\rho\lambda}\left(\delta R\right)^i_{~\rho j \lambda} &=& \frac{1}{a^4}\Bigl( -3{\cal H}'\left( E''\right)^i_{~j}
- 2{\cal H}'{\cal H} \left( E'\right)^i_{~j} \nonumber \\
&& + 6\left( {\cal H}^2 + K\right) {\cal H} \left( E '\right)^i_{~j}
-\left( {\cal H}' + 2\left( {\cal H}^2+K\right)\right)\left( 2{\cal H}^2+\Delta\right)E^i_{~j}\Bigr)~,
\eeq
while since only tensor perturbations are being considered the remaining components vanish.

Calculating the terms in (\ref{eq:EoM_pert}) that involve covariant derivatives is rather more involved
however it is easily achieved by expanding out the covariant derivatives in terms of the metric and
its derivatives. Here we quote the results, with the (tedious) calculation given in Appendix~B.
\beq\label{eq:curvs_pert_2}
\delta \left( R^{;\rho}_{~~;\rho}\right) &=&
  \delta \left( R^{;0}_{~~;0}\right) = \delta \left( R^{;0}_{~~;i}\right) = 0~, \nonumber \\
\delta \left( R^{;i}_{~~;j}\right) &=& \frac{-1}{a^2}  \bar{R}' \left( E'\right)^i_{~j}~, \nonumber \\
\delta\left( R^{0 ~;\rho}_{~0~;\rho}\right) &=& \delta\left( R^{0 ~;\rho}_{~i~;\rho}\right)=0~, \nonumber \\
\delta\left( R^{i ~;\rho}_{~j~;\rho}\right) &=& \frac{8}{a^4} {\cal H} \left( {\cal H}' - {\cal H}^2 - K\right)
\left( E'\right)^i_{~j} \nonumber \\
&& + \frac{1}{a^2}{\cal D}
  \left( \frac{1}{a^2}\left( -\left( E''\right)^i_{~j} - 2{\cal H}\left( E'\right)^i_{~j} + \left( \Delta - K\right)
E^i_{~j}\right)\right)~,
\eeq
where in the last equation, the differential operator ${\cal D}$ is defined as
\be
 {\cal D} \equiv\left( -\frac{\partial^2}{\partial \eta^2} - 2{\cal H}\partial_\eta + 2{\cal H}^2 + \Delta\right)~.
\ee

Using the expressions given in (\ref{eq:curvs_pert_2}) one readily finds that the evolution equation
for tensor mode perturbations, (\ref{eq:EoM_pert}), becomes
\beq\label{eq:EoM_pert_eta}
 \delta H^i_{~j} = \frac{1}{a^4} \Bigl[ - \alpha \left( E''''\right)^i_{~j} + \left( \alpha_1 
 +2\alpha \Delta \right)\left( E''\right)^i_{~j} + \alpha_2 \left( E'\right)^i_{~j} \nonumber \\
  +\left[ \left( \Delta - K\right)\left( \alpha_3 -\alpha\Delta \right) + \alpha_4 - 2\alpha\left( {\cal H}'
+2{\cal H}^2 + 2K \right)\Delta \right] E^i_{~j} \Bigr]~, \nonumber \\
\eeq
where we have defined the $\eta$ dependent functions,
\beq
 \alpha_1 &\equiv& -4{\cal H}' \left( 2\alpha + 3\beta\right) + 4{\cal H}^2\left( \alpha-3\beta\right)
 - 3K\left( \alpha + 4\beta\right) -\frac{\gamma a^2}{\kappa^2}~, \nonumber \\
 \alpha_2 &\equiv& 4{\cal H}''\left( \alpha-3\beta\right) + 8\alpha{\cal H}^3 + 6\alpha{\cal H}K
 - 4{\cal H}{\cal H}'\left( \alpha+6\beta\right) - \frac{2\gamma a^2 {\cal H}}{\kappa^2}~, \nonumber \\
 \alpha_3 &\equiv& -2{\cal H}'\left( \alpha-6\beta\right) + 12\beta {\cal H}^2 + 2K\left( \alpha+6\beta\right)
 +\frac{\gamma a^2}{\kappa^2}~, \nonumber \\
 \alpha_4 &\equiv& 6\alpha{\cal H}^2K + 4\alpha K{\cal H}' + 6\alpha K^2~.
\eeq

Notice in particular that the degeneracy present in the background equations between $\alpha$ and
$\beta$ is now broken and the particular combination $\alpha-3\beta$ play no special role in the
perturbation equations. This agrees with the results of~\cite{Nelson:2008uy,Nelson:2010rt},
which showed that deviations from standard cosmology for NCG inspired corrections to GR (for
which $\alpha = 3\beta$), are important for perturbations in general and gravitational waves in particular.
Indeed, by restricting our attention to flat Minkowski space-time, i.e., ${\cal H}=0$ and $K=0$, and
taking $\alpha = 3\beta$, 
(\ref{eq:EoM_pert_eta}) reduces to the equation for gravitational radiation derived in~\cite{Nelson:2010rt}.

\section{Tensor Perturbations Against Different Cosmological Backgrounds}\label{sec:tensors_BG}

The general evolution equation given in
(\ref{eq:EoM_pert_eta}) can be simplified by considering specific forms of background evolution. For example,
consider $a\left( \eta\right) = a_{\rm i}\eta_{\rm i}^{-\nu} \eta ^\nu$ for some constants $\nu$, $a_{\rm i}$
and $\eta_{\rm i}$.  Such power law behaviour is 
typical in cosmologies dominated by a single matter fluid {\it in GR} e.g.\ $\nu=2$ corresponds to the matter dominated
era, while $\nu=1$ gives the radiation dominated era, with $a_{\rm i}$ and $\eta_{\rm i}$ being
the scale factor and conformal time at the beginning of these eras respectively (see Section~\ref{sec:BG}).
One can easily extend this to include several matter
fields and hence allow for (for example) the transition from the radiation to matter dominated eras, however for simplicity
we focus on single fluid systems. Here we have kept $\alpha$ and $\beta$ general, however it must be remembered
that if (for example) $a\left( \eta\right)\propto \eta$ is to represent the expansion during a radiation
dominated era, then we must satisfy (\ref{eq:rad_con}).

Decomposing all spatial functions into eigenfunctions, $Q_{\bf k}$ of the spatial Laplacian i.e.\
$\Delta Q_{\bf k} = -|{\bf k}|^2 Q_{\bf k}$~\footnote{For the spatially flat, $K=0$ case this is just the usual
Fourier decomposition. For $K=1$ the eigenvalues take discrete values, $|{\bf k}|^2 = l\left( l+2\right)$ for
$l\in \mathbb{Z}$ and for $K=-1$ the eigenvalues are bounded below by $|{\bf k}|^2 >1$~\cite{durrer_book}.} so that
in particular $\Delta E^i_{~j} = -|{\bf k}|^2 E^i_{~j}$, one finds,
\beq\label{eq:prop_pow}
\delta H^i_{~j}\Big|_{\rm power law} = \left( \frac{|{\bf k}|}{a_{\rm i}}\right)^4\left( \frac{x_{\rm i}}{x}\right)^{4\nu}
 \Biggl[ -\alpha \frac{{\rm d}^4 E^i_{~j}}{{\rm d}x^4 }
 + \left( -\beta_1 +\frac{\beta_2}{x^2} - {\cal K}\left( \frac{x}{x_{\rm i}} \right)^{2\nu}\right)
  \frac{{\rm d}^2 E^i_{~j}}{{\rm d} x^2} \nonumber \\
 + \left( 6\alpha\nu\frac{K}{|{\bf k}|^2} +  \frac{\beta_3}{x^2} - 2\nu{\cal K} \left( \frac{x}{x_{\rm i}}\right)^{2\nu}\right)
\frac{1}{x}  \frac{{\rm d}E^i_{~j}}{{\rm d} x} \nonumber \\
 + \left( \beta_4 + \frac{\beta_5}{x^2} + x^2\beta_6 - {\cal K} \left( \frac{x}{x_{\rm i}}\right)^2
\left( \frac{K}{|{\bf k}|^2} + 1 \right)\right) E^i_{~j} \Biggr]~,
\eeq
where
\beq\label{eq:betas}
 \beta_1 &=& 2\alpha + 3\left( \alpha+4\beta\right)\frac{K}{|{\bf k}|^2}~, \nonumber \\
 \beta_2 &=& 4\nu\left( 2\alpha+3\beta + \nu\left( \alpha - 3\beta\right) \right)~, \nonumber \\
 \beta_3 &=& 8\nu^3\alpha + 4\nu^2\left( \alpha+6\beta\right) + 8 \nu\left( \alpha - 3\beta\right)~, \nonumber \\
 \beta_4 &=& 4\left( \alpha-3\beta\right)\left( \frac{K}{|{\bf k}|^2}\right)^2 + 4\alpha \left( \frac{K}{|{\bf k}|^2}
\right) - \alpha - 2\nu\left( \alpha - 6\beta\right) - 12\nu^2\beta~, \nonumber \\
 \beta _5 &=& 6 \nu \left( \nu\left( \alpha-2\beta\right) - \alpha + 2\beta\right) \left( \frac{K}{|{\bf k}|^2}\right)
 + 2\alpha\nu\left( 2\nu-1\right)~, \nonumber \\
 \beta_6 &=&  - 3\left( \alpha-4\beta\right) \left( \frac{K}{|{\bf k}|^2}\right)~,
\eeq
and we have changed variables to $x=\eta|{\bf k}|$ and defined the (dimensionless) variables
$x_{\rm i} \equiv |{\bf k}|\eta_{\rm i}$ and 
\be\label{eq:K}
 {\cal K} \equiv \frac{\gamma}{\kappa^2} \left( \frac{a_{\rm i}}{|{\bf k}|}\right)^2~.
\ee
Note that if ${\cal K} \gg \alpha, \beta$, then (\ref{eq:prop_pow}) approximates the usual
GR evolution equation for tensor modes (e.g.~\cite{durrer_book}), as it should. In particular
then the evolution of extremely large scale modes (i.e.\ modes for which $|{\bf k}| \ll 1$) is
precisely that of GR.  Of course, here one has to be careful, since the GR limit of
Eq.~(\ref{eq:prop_pow}) reduces the order of the perturbation equation and hence can introduce
strong instabilities. One can, however check that {\it if} the solution matches the GR solution initially, then
it will continue to do so throughout the evolution. Furthermore, one can estimate the magnitude of the
effect of small deviations to the GR solution, by evaluating the higher derivative terms appearing in
Eq.~(\ref{eq:prop_pow}) for the (approximate) GR solutions. Doing this one finds that the corrections
seem to remain small during the evolution of the modes, however such evidence does not remove the possibility
of their being a growing mode to the corrections to the GR solutions, even in the ${\cal K} \gg \alpha, \beta$ limit.
However, if such an instability were present, it would likely involve a significant growth in the 
amplitude of $\delta H^i_{~j}$  and hence would result in even stronger constraints than those presented below.

Of particular interest is the evolution of tensor modes in a flat ($K=0$) universe, during the radiation era.
Recall for such a radiation dominated universe we have $\nu \approx 1$ (see the discussion below
(\ref{eq:rad})). One immediately sees that for $\nu=1$ the coefficients
given in (\ref{eq:betas}) are independent of $\beta$. Hence the propagation of
tensor modes during the radiation era are {\it unaffected} by the presence of $R^2$ corrections,
for a flat universe, regardless of the strength of the coupling $\beta$.
Assuming that the universe is exactly flat, the evolution during the radiation era ($\nu=1$)
for a general $\alpha$ is given by,
\beq\label{eq:rad_evol}
 \delta H^i_{~j} = \alpha \left(\frac{|{\bf k}|}{a_{\rm i}}\right)^4 \left( \frac{x_{\rm i}}{x}\right)^4
\Biggl[ -\frac{{\rm d}^4E^i_{~j}}{{\rm d}x^4} + \left( \frac{12}{x^2} - 2 - \frac{\cal K}{\alpha}\left( \frac{x}{x_{\rm i}}
\right)^2\right) \frac{{\rm d}^2 E^i_{~j}}{{\rm d} x^2} \nonumber \\
  \left( \frac{20}{x^2} - \frac{2{\cal K}}{\alpha} \left( \frac{x}{x_{\rm i}}\right)^2\right)
\frac{1}{x} \frac{{\rm d}E^i_{~j}}{{\rm d}x} + \left( \frac{2}{x^2} - 3 - \frac{{\cal K}}{\alpha}
\left( \frac{x}{x_{\rm i}}\right)^2\right)E^i_{~j} \Biggr]~.
\eeq
Hence, the evolution will be well approximated by that of GR when
\be
 \frac{\cal K}{\alpha} \left( \frac{x}{x_{\rm i}}\right)^2 \gg 1~,
\ee
which is true at late times, i.e., large $x$, (recall however the comments below Eq.~(\ref{eq:K})).

If we consider the evolution of perturbations to a flat inflationary era i.e.\ with 
$K=0$ and $\nu=-1$, and consider {\it only} $R^2$ corrections i.e.\ $\alpha=0$,
we find that (\ref{eq:prop_pow}) factorises into
\beq\label{eq:prop_inf}
 \delta H^i_{~j} \Big|_{\nu=-1,~ \alpha=0,~ K=0} = \nonumber \\
  -\left( \frac{|{\bf k}|}{a}\right)^4 \left( \frac{x}{x_{\rm i}}\right)^2
\left( \frac{24\beta}{x^2_{\rm i}} + {\cal K}\right)
\left( \frac{{\rm d}^2 E^i_{~j}}{{\rm d}x^2} - \frac{2}{x} \frac{{\rm d}E^i_{~j}}{{\rm d}x}
+E^i_{~j}\left( x\right)\right)~. 
\eeq
The differential equation in parenthesis is precisely that of General Relativity,
thus, in the absence of a source, the evolution of tensor mode perturbations
during inflation (with $K=0$), is unaffected by the presence of $R^2$ corrections (provided there
are no $R^{\mu\nu}R_{\mu\nu}$ corrections). Note
however that although the evolution of these modes is independent of such 
corrections, their {\it generation} will not be. As can be seen from
(\ref{eq:prop_inf}) the effect of the $R^2$ correction is to introduce 
an effective, Newtonian constant
\be
 a^2_{\rm i}\frac{\gamma}{\kappa^2} \rightarrow a^2_{\rm i}\frac{\gamma}{\kappa^2} + 24
\frac{\beta}{\eta_{\rm i}^2}~.
\ee
In particular there
is a critical value at $24\beta\kappa^2 = -\gamma\left( a_{\rm i}
\eta_{\rm i}\right)^2$
at which the evolution becomes ill defined. This is precisely the same divergence found
for the background evolution in Section~(\ref{sec:ds_bg}). The vanishing of the
prefactor in (\ref{eq:prop_inf}) makes the evolution arbitrarily sensitive to
any deviations for this equation of motion (i.e.\ the evolution is unstable), which is
to be expected for perturbations about such a pathological background cosmology.

We mention in passing that for the specific ratio $\alpha = 24\beta$ and $K=0$,
(\ref{eq:prop_pow}) can be exactly solved. The resulting expression,
involving hypergeometric functions, is not, however, particularly 
illuminating.

\section{Constraining modifications to GR}\label{sec:constraints}
As mentioned in Section~\ref{sec:1}, local tests of GR, particularly from the
perihelion precession of Mercury~\cite{Stelle:1977ry}  and the production of
gravitational waves from binaries~\cite{Nelson:2010ru}, provide only very weak constraints,
typically of the order of  $\left(\alpha-3\beta\right) < 10^{70}$.
Cosmology has the advantage over such test, both because it is possible to reach higher curvature
regimes in the early universe and also because we naturally have very long evolution times, during
which even small deviations from GR can become significant. As we found in Section~\ref{sec:BG}, in order
for a slowly rolling scalar field to have sustained the exponential expansion of inflation, in the usual
way, we require
\be\label{eq:constraint_inf}
 \left( \frac{|\alpha-3\beta|}{\gamma \kappa^{-2}}\right) \left(\frac{1}{a_{\rm inf}\eta_{\inf}}\right)^2
\ll 1~,
\ee
where recall $a_{\rm inf}$ and $\eta_{\rm inf}$ are the scale factor and conformal time
at the beginning of inflation. Note that $\eta$ is conformal time, so $a_{\rm inf}
\eta_{\rm inf} = t_{\inf}$, where $t_{\rm inf}$ is the {\rm cosmic} time at the beginning
of inflation. Taking inflation to have occurred $t\approx t_0\times 10^{-40} \approx 10^{-22}
{\rm sec}$~\cite{kolb_turner},
where $t_0$ is the current {\it cosmic} time (taken to be approximately $10^{18} {\rm sec}$),
we see that (\ref{eq:constraint_inf}) becomes,
\be
 \left( \frac{|\alpha-3\beta|}{\gamma\kappa^{-2}}\right) \ll 10^{-80}t_0^2~,
\ee
or, taking $\gamma\kappa^{-2} = \left( 16 \pi G\right)^{-1}$ and using the experimental value of Newton's
constant ($G\approx 2.88 \times 10^{-87}{\rm sec}^2$ in these units),
\be
 |\alpha - 3\beta| \ll 6.7 \times 10 ^{40}~.
\ee
While this constraint may still seem rather weak it is an improvement by some {\it thirty orders
of magnitude} over the previous best astrophysical constraints (and is twenty orders of magnitude
better than laboratory tests).

One may be concerned that this constraint is produced by assuming that inflation is produced by
a slowly rolling scalar field. Maybe there is some other mechanism for producing the exponential
expansion required to solve the horizon, curvature and monopole problems of the standard big-bang theory?
Indeed, there are examples of higher derivative theories driving inflation directly~\cite{Starobinsky:2001xq,Nelson:2009wr}.
Or perhaps there is an alternative to inflation that does not require such exponential expansion?
Certainly, since we have yet to derive the consequences of this general $4^{\rm th}$ order
theory on the growth (or otherwise)
of scalar perturbations, it is premature to suggest that such an expansion phase 
would produce the scale invariant perturbation spectrum required by the CMB.

To avoid these difficulties let us take the more conservative point of view and assume that
the dynamics of the
universe closely match those expected from GR only after some time, $t_{c}$, in the radiation era.
Certainly the success of the CMB tells us that GR is a good approximation significantly before
decoupling (i.e.\ before the last scattering surface). From (\ref{eq:rad_con}) we then have
\be
 \left( \frac{|\alpha-3\beta|}{\gamma\kappa^{-2}}\right) \frac{1}{t_{\rm r}^2}\left( 
\frac{a_{\rm r}}{a_{\rm c}}\right)^2 \ll 1~,
\ee
where $t_{\rm r}$ is the {\it cosmic} time at the beginning of the radiation era and
$a_{\rm r}$ and $a_{\rm c}$ are the scale factors at cosmic times $t_{\rm r}$ and $t_{\rm c}$
respectively. Being extremely conservative, let us assume that the universe was
radiation dominated at energy scales below electro-weak unification ($t_{\rm EW} \sim 10^{-12}{\rm sec}$)
 $t_{\rm r}\sim
10^{-28}t_0$\footnote{ Here we are estimating the cosmic time at which this occurs
by the GR value. Since we are assuming that GR is {\it not} a good approximation
until we reach times beyond $t_{\rm c} > t_{\rm r}$, this is not strictly consistent.
However we are only estimating orders of magnitude, for which this should be sufficient.}.
 Further let us assume that
the universe is well approximated by that of GR at the scales on which Big Bang
Nucleosynthesis takes place, $t_{\rm c} \approx 10^{-18}t_0$~\cite{kolb_turner}. This
assumptions gives
\be
 \frac{|\alpha-3\beta|}{\gamma\kappa^{-2}} \ll 10^{-50} t_0^2~,
\ee
or, putting in the current age of the universe and Newton's constant,
$|\alpha - 3\beta| \ll 10^{70}$. Thus even with these extremely conservative
estimates, the constraint is the same as that from gravitational waves and
solar system tests. 

While the estimates above can provide a significant improvement on existing constraints,
in the background equations there is a degeneracy in the parameters, in that we
can restrict only the combination $\alpha - 3\beta$. In order to place independent
constraints on each parameter, we need to consider perturbations around the background
cosmology. Using the results of Section~\ref{sec:tensors_BG} we see that,
during the radiation era, the evolution of
the cosmological gravitational waves will deviate significantly from the predictions
of GR unless,
\be\label{eq:rad_constraint}
 \frac{\cal K}{\alpha}\left(\frac{x}{x_{\rm i}}\right)^2 \gg 1~.
\ee
As a specific example, consider a mode that crosses the Horizon at matter-radiation
equality\footnote{Close to matter-radiation equality, one should use a two fluid approximation,
however we are concerned only with rough estimates and hence neglect this complication.},
 which occurs at $t_{\rm eq} \sim 10^{12}{\rm sec}$,
and again conservatively consider the beginning of the radiation dominated era to be
at the electro-weak scale. One then finds that 
(\ref{eq:rad_constraint}) reduces to,
\be\label{eq:tensor_constraint}
 \alpha \ll 6.7\times 10^{70}~,
\ee
again comparable to the current best constraints on such parameters. As is clear from
(\ref{eq:rad_constraint}), this constraint can be improved in two ways;
by decreasing $x/x_{\rm i}$ i.e. considering the modes close to the beginning of the
radiation era or by decreasing ${\cal K}$. Recall from (\ref{eq:K}) that ${\cal K} \propto |{\bf k}|^{-2}$,
thus the constraint can be improved by considering {\it smaller} wavelength gravitational
waves (i.e.\ larger $|{\bf k}|$).
 The fact that these deviations from GR become more significant at small scales 
might have been expected because  of the fact that these modes are ghosts~\cite{Clunan:2009er}.

Taking the point of view that this theory is only an effective theory, good down to some scale,
after which some more complete theory will dictate the evolution of these modes, we 
require only that the deviation for GR be not significant (i.e.\ that the ghosts are not
apparent) on observational scales. The use of the word {\it observational} here deserves some
comment, since we have yet to observe cosmological gravitational waves at any scale! However
the deviations from GR will typically make the tensor perturbations grow (as they represent
an instability). If such growth had occurred and been significant, we would have observed their
contribution to the galaxy power spectrum and the CMB. A quantitative constraint would require
an in-depth analysis of the exact evolution given in (\ref{eq:rad_evol}) and an understanding
of the rate of the growth of the modes that can be expected. However in order to get an
estimate on the order of magnitude of the constraint, we will simply require that 
the deviations from GR be small at the smallest observable scales at which such cosmological
tensor modes could have been observed i.e.\ for the first modes that entered the horizon 
that have been observed in the galaxy power spectra.

Data from Lyman alpha forests~\cite{Tegmark:2001jh,Tegmark:2002cy} provide us with a detailed power
spectrum of (scalar) density perturbations that matches the expectations of GR down to 
scales $|{\bf k}| \sim \left( 10^{2} \-- 10^3\right) |{\bf k}|_{\rm equality}$, where
$|{\bf k}|_{\rm equality}$ is the scale at matter radiation equality (which produces
the characteristic peak in the power spectrum). Thus at such scales the constraint is
improved by a factor of $10^{4} \-- 10^{6}$ over (\ref{eq:tensor_constraint}).
This improvement would be further increased by considering smaller scales or, of 
course, by direct measurement of the cosmological gravitational wave background.

\section{Conclusions}\label{sec:conc}

We have shown that it is possible to approximately reproduce the key epochs of standard cosmology
within a generally modified $4^{\rm th}$ order gravity theory. More precisely we
demonstrated that {\it only} a correction to Einstein-Hilbert gravity of the form
$C^{\mu\nu\rho\gamma}C_{\mu\nu\rho\gamma}$ introduces no deviation from standard
background cosmology, while any other correction can approximate the standard
background evolution, provided it is initially `close' to such a correction (see
Section~\ref{sec:BG} for a more precise statement). Furthermore by requiring that
the deviations from GR be small at certain key eras (such as during inflation
and Big Bang Nucleosynthesis) we can quantify how `close' general corrections need to
be in order to be consistent. Even taking the conservative point of view that
radiation domination began only after the electroweak scale, such constraints
are comparable to the best current restrictions on the theory. Further requiring
that inflation be sourced by a fluid with $\omega \approx -1$ allowed us to improve
this constraint {\it by thirty orders of magnitude}! The appearance of such 
large constraints is not uncommon within cosmology. Consider for example the 
spatial curvature within non-inflationary cosmology. If the energy density associated
with such curvature vanishes to one part in $10$ today, it would have to vanish to one part in
$10^{60}$ close to the Grand Unification scale. Indeed this was one of the original motivations for
inflation. Similarly requiring that the higher order curvature corrections be small
today is a {\it much} weaker constraint than requiring they be small in the past. The
earlier we can examine the consequences of such corrections (such as during 
the radiation era or inflation) the more the constraints can be improved.

At the background level we are able only to constrain the parameter $\alpha-3\beta$
($\alpha=3\beta$ corresponds to the $C^{\mu\nu\rho\gamma}C_{\mu\nu\rho\gamma}$ correction
of Non-commutative Geometry)
and not the individual coefficients. In order to break this degeneracy we consider
the evolution of perturbations against the homogeneous and isotropic background.
For comparison with observations, one would like to examine the scalar perturbations
of such a theory, which can be directly related to the galaxy power spectrum,
weak lensing surveys, and the CMB. However the equations governing the evolution
of scalar perturbation are extremely involved in a general $4^{\rm th}$ order
gravity theory, making their use difficult. Instead we focused on the evolution of
tensor modes (cosmological gravitational waves), which are technically simpler to 
calculate. While they have yet to be observed, they have the additional advantage of
not coupling directly to matter (in the absence of anisotropic stress). Thus 
they are independent of any corrections to the matter action and are direct probes 
of the underlying geometry of the problem.

Without direct observational evidence of such tensor mode perturbations, we 
considered the types of gravitational theories for which the evolution is 
approximately that of GR. The fact that such theories will typically lead
to instabilities of perturbations, suggests that deviations from GR will
include a strong growth of the amplitude of the perturbations. Had this occurred
the perturbations would have been observed and hence our constraint is equivalent to 
ensuring that such instabilities occur at scales that are not (yet) observable.

We showed that the presence of $R^2$ corrections to GR does not affect the
evolution of tensor mode perturbations {\it in a flat radiation dominated 
universe}, while if {\it only} $R^2$ corrections are present, there
is also no effect on the evolution, during the inflation era (although
one can expect an affect on their production). Requiring that corrections
be small for a specific mode and a particular scale (within the radiation
era) hence constrains the coefficient of $R^{\mu\nu}R_{\mu\nu}$ corrections
and breaks the degeneracy present in the background equations.

By considering modes that have contributed to the measured (scalar) power
spectrum\footnote{To avoid confusion we emphasise again, the tensor modes
have {\it not} been measured. We consider these modes only because their
scalar counterparts have been observed and quantified and if significant
enhancement of the tensor modes had occur they would have contributed to the
observed power spectrum at these scales.} we can estimate the constraint on $R^{\mu\nu}R_{\mu\nu}$
corrections, finding it to be improved by factor of $ 10^{4} \-- 10^{6}$ 
over current best restrictions. Thus, if inflation is to have occurred in the usual manner, we require
\be
 \alpha - 3\beta \ll {\cal O}\left( 10^{40}\right)~, ~~~~ \alpha \ll {\cal O}
\left( 10^{65}\right)~,
\ee
which is a remarkable improvement on current constraints which give $\alpha,~\beta
\ll {\cal O}\left(10^{70}\right)$.

Finally it is important to note that
the presence of ghost in theories with such $4^{\rm th}$ order corrections
is well known~\cite{Stelle:1977ry} and clearly indicates that the theory cannot be a
fundamental theory (and certainly cannot be quantised in any standard way).
Thus one should consider such corrections as an effective theory, which is
valid in some range of scales and replaced by some more complete
theory outside that range. For example, such corrections occur in the
asymptotic expansion of certain Non-commutative Geometry theories~\cite{Chamseddine:2010ud},
(in that case $\alpha =3\beta$) and beyond a specific scale the space-time
is no longer even approximately commutative. Here we have shown that the
 presence of ghosts is not
felt at the level of background cosmology, however it is crucial for the
evolution of perturbations and the constraints produced here essentially 
ensure that the presence of ghosts would not effect observable scales.
One expects that higher (i.e.\ $6^{\rm th}$ and higher~\cite{Gottlober:1989ww,Hwang:1991}) order 
corrections to~(\ref{eq:action}) would become significant some scale.
The presence of ghosts indicate the scale at which these higher
order terms would need to become significant and essentially mark
the scale at which the ($4^{\rm th}$ order) 
effective theory begins to break down. Thus the constraints produced here
can be viewed as determining the scale at which (\ref{eq:action}) is a valid 
effective theory. It is important to note however, that here we have assumed
a perturbative expansion and it is possible that non-perturbative effects
might be significant.

\section*{Acknowledgments}
I would like to thank N. Johnson-McDaniel, M. Sakellariadou and A. Tsobanjan
for useful comments and suggestions.
This work is supported in part by the NSF grant PHY0854743,
the George A. and Margaret M. Downsbrough Endowment and the Eberly
research funds of Penn State.

\appendix\label{app}
\section*{Appendix A}
\setcounter{section}{1}

In this appendix we derive the time-time component and trace of the tensor $H^\mu_{~\nu}$ given in
(\ref{eq:EoM}), for the background FRW metric, (\ref{eq:metric_bg}). First note that the
non-zero Christoffel symbols associated with this metric are,
\beq\label{eq:christ_bg}
 \bar{\Gamma}^0_{00} = {\cal H}~, ~~~~ \bar{\Gamma}^i_{j0} = {\cal H}\delta ^i_{~j}~, ~~~~ \bar{\Gamma}^0_{ij}
= {\cal H}\gamma_{ij}~, ~~~~ \bar{\Gamma}^k_{ij} =~ ^{(3)}\bar{\Gamma}^k_{ik}~,
\eeq
where $^{(3)}\bar{\Gamma}^k_{ij}$ is the three dimensional Christoffel symbol calculated from the
spatial metric $\gamma_{ij}$. In order to evaluate the components of (\ref{eq:EoM}) we need,
\beq
\bar{R}^{\mu~;\rho}_{~\nu~;\rho} &=& \frac{-1}{a^2} \left( \bar{f}^\mu_{~\nu0,0} + \bar{\Gamma}^\mu_{0\alpha}
\bar{f}^\alpha_{~\nu0} - \bar{\Gamma}^\alpha_{\nu 0}\bar{f}^\mu_{~\alpha 0} -\bar{\Gamma}^\alpha_{~00} 
\bar{f}^\mu_{~\nu\alpha}\right) \nonumber \\
&& + \frac{\gamma^{ij}}{a^2}\left( \bar{\Gamma}^\mu_{i\alpha}f^\alpha_{~\nu j}
-\bar{\Gamma}^\alpha_{\nu i} \bar{f}^\mu_{~\alpha j} - \bar{\Gamma}^\alpha_{ji} \bar{f}^\mu_{~\nu\alpha}\right)~,
\eeq
where we have defined $\bar{f}^\mu_{~\nu\alpha} \equiv \bar{R}^\mu_{~\nu;\alpha}$, which has non-zero 
components,
\beq
\bar{f}^0_{~00} &=& \frac{3}{a^2} \left( 2{\cal H} {\cal H}'- {\cal H}''\right)~, \nonumber \\
\bar{f}^0_{~ij} &=& \frac{2{\cal H}}{a^2} \gamma_{ij} \left( {\cal H}' - {\cal H}^2 - K\right)~, \nonumber \\
\bar{f}^i_{~0j} &=& \frac{-2{\cal H}}{a^2} \delta^i_{~j} \left( {\cal H}'-{\cal H}^2 - K\right)~, \nonumber \\
\bar{f}^i_{~j0} &=& -\delta^i_{~j} \left( \frac{1}{a^2} \left( {\cal H}' + 2\left( {\cal H}^2 + K\right)\right)\right)'~.
\eeq
Thus we find,
\beq
 \bar{R}^{0~;\rho}_{~0~;\rho} &=& \frac{-3}{a^4}\left( 2\left({\cal H}'\right)^2 + 2{\cal H}{\cal H}''
-{\cal H}''' + 4{\cal H}^2{\cal H}' - 4{\cal H}^4 - 4{\cal H}^2K \right)~,\nonumber \\
 \bar{R}^{\gamma~;\rho}_{~\gamma~;\rho}&=& \frac{6}{a^4} \left( {\cal H}''' - 2{\cal H}'\left( K+3{\cal H}^2\right)\right)~.
\eeq

In addition we need $\bar{R}^{;0}_{~;0}$ and $\bar{R}^{;\rho}_{~;\rho}$ which are,
\beq
\bar{R}^{;0}_{~;0} &=& \frac{6}{a^4} \left( -6{\cal H}^2{\cal H}' -3{\cal H}{\cal H}''
+6{\cal H}^2\left( {\cal H}^2+K\right) - 2{\cal H}'K + {\cal H}'''\right)~, \nonumber \\
\bar{R}^{;\rho}_{~;\rho} &=& \frac{6}{a^4} \left( {\cal H}''' - 6{\cal H}^2{\cal H}' - 2{\cal H}K\right)~.
\eeq

Using the expressions given in (\ref{eq:curvs}) one directly finds the remaining terms in $H^0_{~0}$ and
$H^\rho_{~\rho}$, eventually giving,
\beq
 \bar{H}^0_{~0} &=& \frac{6\left( \alpha - 3\beta\right)}{a^4} \left[ -2{\cal H} \left( {\cal H}'' + 2{\cal H}
{\cal H}'\right) + 3{\cal H}^4 + 2{\cal H}^2K + \left( {\cal H}'\right)^2 -K^2\right] \nonumber \\
&& + \frac{3\gamma\kappa^{-2}}{a^2}\left( {\cal H}^2+K\right) = \frac{\rho}{2}~, \\
 \bar{H}^{\rho}_{~\rho} &=& \frac{-2\left( \alpha-3\beta\right)}{a^4}
\left[ {\cal H}''' - 6{\cal H}^2{\cal H}' - 2{\cal H}'K\right] \nonumber \\
&& + \frac{\gamma\kappa^{-2}}{a^2}\left( {\cal H}' +{\cal H}^2
+K\right) = \frac{1}{12}\left( \rho - 3P\right)~,
\eeq
which are the equations given in (\ref{eq:Fried_ray_BG}).

\appendix
\section*{Appendix B}\label{app2}
\setcounter{section}{2}

Here we calculate the perturbations to $R^{;\rho}_{~~;\rho}$, $R^{;\mu}_{~~;\nu}$ and $R^{\mu~;\rho}_{~\nu~;\rho}$.
In fact, since $R^{;\rho}_{~~;\rho}$ is a scalar and we are considering only tensor perturbations,
we know that its perturbation will vanish~\footnote{Similarly $R^{;0}_{~~;0}$, $R^{;0}_{~~;i}$, $R^{0~;\rho}_{~0~;\rho}$
and $R^{0~;\rho}_{~i~;\rho}$ will vanish as these are either scalars or vectors.}, however for pedagogical reason we explicitly
calculate it here.

Expanding the covariant derivatives in terms of (derivatives of) the metric we have,
\be
  R^{;\rho}_{~~;\rho} = g^{\alpha\beta}\left( \partial_\alpha \partial_\beta R - \Gamma ^\gamma_{~\beta \alpha}
\partial_\gamma R\right)~,
\ee
where $\Gamma^\gamma_{~\beta\alpha}$ is the Christoffel symbol. Perturbing this gives,
\beq\label{eq:christ}
 \delta\left(  R^{;\rho}_{~~;\rho}\right) &=& \delta g^{\alpha\beta}\left( \partial_\alpha \partial_\beta \bar{R} - 
\bar{\Gamma} ^\gamma_{~\beta \alpha} \partial_\gamma \bar{R}\right)  \nonumber \\
&&+ \bar{g}^{\alpha\beta}\left( \partial_\alpha \partial_\beta \delta R - \left(\delta \Gamma\right) ^\gamma_{~\beta \alpha}
 \partial_\gamma \bar{R}- \bar{ \Gamma} ^\gamma_{~\beta \alpha}
 \partial_\gamma \delta R\right)~.
\eeq
To calculate this note that $\delta R=0$ and the non-zero components of the background
Christoffel symbols are given in (\ref{eq:christ_bg}) and the perturbed versions are,
\beq
&& \delta \Gamma^i_{~0j} = \left( E'\right)^i_{~j}~, ~~~~ \delta \Gamma^0_{~ij} = 2{\cal H}E_{ij}
+E'_{ij}~,~~~~ \delta \Gamma^i_{~jk} = E^i_{~j|k} + E^i_{~k|j} - E^{~~|i}_{kj}~.\nonumber \\
\eeq
Using this and the expressions in (\ref{eq:curvs}) and
(\ref{eq:metric_perts}) one finds,
\beq
 \delta\left(  R^{;\rho}_{~~;\rho}\right) &=& - \delta g^{ij} \bar{\Gamma}^0_{~ij} \bar{R}'
 - \bar{g}^{ij}\left( \delta \Gamma\right)^0_{~ij} \bar{R}'~, \nonumber \\
&=& \frac{2}{a^2} E^{ij} {\cal H} \gamma_{ij} \bar{R}' - \frac{1}{a^2} \gamma^{ij} \left(
2{\cal H} E_{ij} + E'_{ij}\right)\bar{R}'~,
\eeq
which, as expected, vanishes because our perturbations are traceless i.e.\ $E^{ij}\gamma_{ij}=0$.

One can perform a similar expansion of the covariant derivatives in $R^{;\mu}_{~;\nu}$ to find,
\be
 R^{;\mu}_{~;\nu} = \left( g^{\mu\alpha} R_{,\alpha}\right)_{,\nu} + \Gamma^\mu_{~\nu\beta}
g^{\beta \alpha} R_{,\alpha}~.
\ee
Perturbing this one finds (here we only explicitly calculate the non-zero components),
\beq
 \delta \left( R^{;i}_{~;j} \right) &=& \delta\Gamma^i_{~j0} \bar{g}^{0\alpha} \bar{R}_{,\alpha}
+ \delta \Gamma^i_{~jk} \bar{g} ^{k\alpha} \bar{R}_{,\alpha} + \bar{\Gamma}^i_{~j0} \delta g^{0\alpha}
\bar{R}_{,\alpha} + \bar{\Gamma}^i_{~jk} \delta g^{k\alpha} \bar{R}_{,\alpha}~, \nonumber \\
&=& \frac{-1}{a^2} \bar{R}'\left( E'\right)^i_{~j}~,
\eeq
where we have used (\ref{eq:christ}) and (\ref{eq:metric_perts}). This is the expression
given in (\ref{eq:curvs_pert_2}).

The final quantity that we want to perturb is $R^{\mu ~;\rho}_{~\nu~;\rho}$.
Again, expanding the covariant derivatives one finds,
\beq\label{eq:box_R_mu_nu}
  R^{\mu ~;\rho}_{~\nu~;\rho} = g^{\gamma \delta}\left( f^\mu_{~\nu\delta,\gamma} 
 +\Gamma^\mu_{~\gamma \alpha} f^\alpha_{~\nu\delta} - \Gamma^\alpha_{~\nu\gamma}
 f^\mu_{~\alpha \delta} - \Gamma^\alpha_{~\delta \gamma} f^\mu_{~\nu\alpha}\right)~,
\eeq
where 
\be
 f^\mu_{~\nu\delta} \equiv R^\mu_{~\nu;\delta} = R^\mu_{~\nu,\delta} +\Gamma^\mu_{~\delta\alpha}
 R^\alpha_{~\nu} - \Gamma^\alpha_{~\delta\nu} R^\mu_{~\alpha}~.
\ee
Perturbing this one finds that the non-zero components of $\bar{f}^\alpha_{~\beta\gamma}$ and
$\delta f^\alpha_{~\beta\gamma}$ are,
\beq
 &&\bar{f}^0_{~00} = \frac{3}{a^2} \left( 2{\cal H}{\cal H}' - {\cal H}''\right)~,~~~~
    \bar{f}^0_{~ij} = \frac{2{\cal H}\gamma_{ij}}{a^2} \left( {\cal H}' - {\cal H}^2 - K\right)~,\nonumber \\
 && \bar{f}^i_{~0j} = \frac{ -2{\cal H} \delta ^i_{~j}}{a^2} \left( {\cal H}' - {\cal H}^2 -K\right)~,~~~~
 \bar{f}^i_{~j0} = -\delta^i_{~j} \left[ \frac{1}{a^2} \left( {\cal H}'+2\left( {\cal H}^2 + K\right)\right)\right]'
 \nonumber \\
 && \delta f^0_{~ij} = \frac{2}{a^2} \left( {\cal H}' - {\cal H}^2 - K\right) \left( 2{\cal H}E_{ij} 
 + E'_{ij}\right)  + {\cal H} {\cal E}_{ij}~, \nonumber \\
&& \delta f^i_{~0j} = \frac{-2}{a^2} \left( {\cal H}' - {\cal H}^2 -K\right) \left( E'\right)^i_{~j}
 -{\cal H}{\cal E}^i_{~j}~,\nonumber \\
&& \delta f^i_{~j0} = \left( {\cal E}'\right)^i_{~j}~, ~~~~ \delta f^i_{~jk} = {\cal E}^i_{~j|k}~,
\eeq
where
\be
 {\cal E}_{ij} \equiv \frac{1}{a^2} \left( - \left(E''\right)_{ij} - 2{\cal H}\left(E'\right)_{ij}
 +\left( \Delta - K\right) E_{ij}\right)~.
\ee
Perturbing (\ref{eq:box_R_mu_nu}) we find
\be
 \delta\left( R^{i ~;\rho}_{~j~;\rho}\right) = \left( T_1\right)^i_{~j}
 + \left( T_2\right)^i_{~j} +  \left( T_3\right)^i_{~j}~,
\ee
where the (components of the) three tensors $T_1$, $T_2$ and $T_3$ are given by,
\beq
 \left( T_1\right)^i_{~j} &\equiv& \delta g^{km} \left( \bar{\Gamma}^i_{~k0} \bar{f} ^0_{~jm} - \bar{\Gamma}^0_{~jk}
\bar{f}^i_{~0m} - \bar{\Gamma}^0_{~mk} \bar{f}^i_{~j0}\right)~, \nonumber \\
&=&\frac{-8}{a^4} {\cal H}^2\left( {\cal H}' - {\cal H}^2 - K\right)E^i_{~j}~, \\
\left( T_2\right)^i_{~j} &\equiv& 
 \bar{g}^{00}\bigl( \delta f^i_{~j0,0} + \delta \Gamma^i_{~0m} \bar{f}^m_{~j0} + \bar{\Gamma}^i_{~0m}
 \delta f^m_{~j0} - \delta \Gamma^m_{~j0} \bar{f}^i_{~m0} 
 \nonumber \\
&& ~~~~~~~~ - \bar{\Gamma}^m_{~j0} \delta f^i_{~m0}- \bar{\Gamma}^0_{~00} \delta f^i_{~j0}\bigr)~, \nonumber \\
 &=& \frac{1}{a^2} \left( \left({\cal E}''\right)^i_{~j} - {\cal H}\left( {\cal E}'\right)^i_{~j} \right)~, \\
 \left( T_3\right)^i_{~j} &\equiv& \bar{g}^{km} \Bigl( \delta f^i_{~jm,k} + \delta \Gamma^i_{~k0} \bar{f}^0_{~jm}
 + \bar{\Gamma}^i_{~k0} \delta f^0_{~jm} + \bar{\Gamma}^i_{~kn} \delta f^n_{~jm} - \delta \Gamma^0_{~jk}
\bar{f}^i_{~0m} \nonumber \\
&&- \bar{\Gamma}^0_{~jk} \delta f^i_{~0m} - \bar{\Gamma}^n_{~jk}\delta f^i_{~nm} - \delta \Gamma^0_{~mk}
\bar{f}^i_{~j0} - \bar{\Gamma}^0_{~mk} \delta f^i_{~j0} - \bar{\Gamma}^n_{~mk} \delta f^i_{~jn}\Bigr)~, \nonumber \\
&=& \frac{1}{a^2} \Bigl( {\cal E}^{i~|k}_{~j~|k} + \frac{8}{a^2}{\cal H}\left( {\cal H}' - {\cal H}^2 - K\right)
\left( \left( E'\right)^i_{~j} + {\cal H} E^i_{~j} \right) + 2{\cal H}^2 {\cal E}^i_{~j} \nonumber \\
 &&~~~~~~~~  - 3{\cal H} \left( {\cal E}'\right)^i_{~j}\Bigr)~.
\eeq
Putting these all together one finds
\be
 \delta\left( R^{i ~;\rho}_{~j~;\rho}\right) = \frac{8}{a^4} {\cal H}\left( {\cal H}' - {\cal H}^2 - K\right)
\left( E'\right)^i_{~j} + \frac{1}{a^2} \left( -\partial_t^2 - 2{\cal H} \partial_t +2{\cal H}^2 
+\Delta\right) {\cal E}^i_{~j}~,
\ee
which is the expression given in (\ref{eq:curvs_pert_2}).

\end{document}